\documentclass[12pt]{article}
\usepackage{amsmath}
\begin{document}
\def\thefootnote{\fnsymbol{footnote}}
\begin{flushright}
KANAZAWA-10-01  \\
May, 2010
\end{flushright}
\vspace{ .7cm}
\vspace*{2cm}
\begin{center}
{\LARGE\bf Enhancement of the annihilation of 
dark matter in a radiative seesaw model}\\
\vspace{1 cm}
{\Large Daijiro Suematsu}\footnote{e-mail:~suematsu@hep.s.kanazawa-u.ac.jp},
{\Large Takashi Toma}\footnote{e-mail:~t-toma@hep.s.kanazawa-u.ac.jp}
{\Large and Tetsuro Yoshida}\footnote{e-mail:~yoshida@hep.s.kanazawa-u.ac.jp}
\vspace*{1cm}\\
{\itshape Institute for Theoretical Physics, Kanazawa University,\\
        Kanazawa 920-1192, Japan}\\
\end{center}
 \vspace*{1cm}
{\Large\bf Abstract}

The radiative seesaw model with an inert doublet has been 
shown to be attractive from a viewpoint of both neutrino 
masses and cold dark matter. However, if we apply this model 
to the explanation of the positron excess in the cosmic ray 
observed by PAMELA, a huge boost factor is required although 
it can be automatically explained that no anti-proton excess 
has been observed there. We consider an extension of the model 
to enhance the thermally averaged annihilation cross section 
without changing the features of the model favored by both 
the neutrino oscillation and the relic abundance of dark matter. 
It is shown that the data of PAMELA and Fermi-LAT can be well 
explained in this extended model. Constraints from gamma ray 
observations are also discussed.  

\def\thefootnote{\alph{footnote}}
\newpage
\section{Introduction}

The existence of dark matter (DM) now gives us a clear motivation 
to investigate physics beyond the standard model (SM).
Although we know its relic abundance in the present universe\cite{wmap}, 
its nature is not clarified except that DM should be cold. 
However, recent observational data on cosmic rays may give us
interesting information on its mass and interactions.
PAMELA has reported positron excess in the cosmic ray at the 6 - 100~GeV
range in comparison with the expected background \cite{pamela}.
However, it has observed no anti-proton excess. 
The preliminary report of Fermi-LAT also suggests 
that the total flux of positrons and electrons is larger than the expected
background at regions of 60 - 1000~GeV \cite{fermi}, 
although any bump shown in the ATIC result \cite{atic} 
is not found in that flux. If we consider these results are 
caused by the decay or
the annihilation of DM but not by astrophysical origins, 
they are expected to give us crucial information on the 
nature of DM.
However, it has been pointed out that there is a difficult problem if we
try to understand these results on the basis of DM physics. 

In case of the DM decay, 
DM life time should be extremely long such as $O(10^{26})$~sec
in order to explain the PAMELA positron excess \cite{decay,fks}.
It is not so easy to answer how such a long lifetime can be naturally
realized in the ordinary models for elementary particles, although
there are several proposals to solve this problem.
In case of the DM annihilation, its relic abundance requires the 
thermally averaged annihilation
cross section $\langle\sigma v\rangle$ to be $O(10^{-26})$~cm$^3$/sec 
at its freeze-out period where the typical DM velocity $v_{\rm DM}$ satisfies
$v_{\rm DM}/c\sim 0.2$. On the other hand, the positron excess 
found in the PAMELA experiment requires
$\langle\sigma v\rangle$ to be $O(10^{-23})$~cm$^3$/sec 
for the DM in our Galaxy where $v_{\rm DM}/c\sim 10^{-3}$ is expected
for the averaged DM velocity. 
This means that there should be some large enhancement introduced as
a boost factor usually, which may be caused by particle physics, or 
astrophysics, or both of them. There have been several analyses on 
this point \cite{annih,mindep}.
In this paper, we focus our attention on the particle physics side and 
study a way to overcome this difficulty in
a model for both neutrino mass and DM.

The radiative seesaw model proposed in \cite{ma} gives an attractive scenario
for the neutrino mass and mixing.
They are explained by new physics at TeV scales in this model. 
The model includes DM candidates
whose stability is guaranteed by a discrete symmetry. It forbids
bare Dirac mass terms of neutrinos and then is related to
the smallness of neutrino masses. 
The model has been studied from various points of view
\cite{phenom,ks,cms,sty}. However, if we apply this model 
to explain the PAMELA data by
the DM annihilation, we confront the boost factor problem, unfortunately. 
In this model the annihilation cross section
has a dominant contribution from its $p$-wave part.
Since the $p$-wave contribution becomes smaller for smaller DM relative
velocity $v$, the situation is much worse than the $s$-wave case. 
In fact, this requires a huge boost factor of $O(10^6)$ to explain the 
PAMELA data on the basis of this model unless there are some 
additional astrophysical effects \cite{cms,sty}.  
On the other hand, we should also remind the reader 
that the model has an interesting 
feature favored by the PAMELA data, that is, the DM can annihilate only 
to leptons. 
Moreover, if we impose constraints on the model from the lepton flavor 
violating processes like $\mu\rightarrow e\gamma$, $e^\pm$ should not be
yielded as the primary final states of the annihilation.
Positrons originated from this DM annihilation are generated through 
the decay of $\mu^+$ and $\tau^\pm$ \cite{sty}. 
Model independent analyses suggest that this feature is again favored 
by both data of PAMELA and Fermi-LAT \cite{mindep}. 
Thus, it is an interesting subject for this model to find some solutions for 
this boost factor problem by extending the model without disturbing these 
nice features of the model.\footnote{One solution has been proposed by
considering the decaying DM in a supersymmetric extension of the model
\cite{fks}.}

In this paper we propose a simple extension of the model, 
which makes the Breit-Wigner enhancement of thermally averaged 
annihilation cross section possible. In that model we show
that both data of PAMELA and Fermi-LAT can be well explained without 
assuming an unknown huge boost factor as long as the mass of a scalar
field is finely tuned.
The enhanced annihilation cross section may also predict the large flux of 
gamma ray which is associated with the DM annihilation.
We discuss the consistency with the data for the diffuse gamma ray
from observations in the EGRET, HESS and Fermi-LAT experiments.    

The following parts of the paper are organized as follows.
In section 2 we fix the radiative seesaw model and discuss how all the
neutrino oscillation data, lepton flavor violating processes and the DM
relic abundance can be consistently explained.
After that we extend the model to enhance the DM annihilation cross section
in the present Galaxy. In section 3 we address the features required 
for the explanation of the data of PAMELA and Fermi-LAT.
We also discuss the consistency between the diffuse gamma ray
flux expected in the model and the present experimental data. 
We summarize the paper in section 4.

\section{Breit-Wigner enhancement}
\subsection{A radiative seesaw model}
The radiative seesaw model considered here is an extension of the SM with 
an inert doublet $\eta$ (an additional SU(2)$_L$ doublet scalar with
$\langle\eta\rangle=0$) 
and three gauge singlet right-handed fermions $N_k~(k=1,2,3)$ \cite{ma}. 
In order to forbid tree-level Dirac masses for neutrinos, we impose  
$Z_2$ symmetry on the model. An odd charge of this $Z_2$ symmetry is
assigned to all of these new fields, although an even charge is assigned 
to all of the SM contents.
Both interaction Lagrangian ${\cal L}_N$ 
relevant to $N_k$ and scalar potential $V$ invariant under the 
imposed symmetry are written as 
\begin{eqnarray}
{\cal L}_N&=& -\left(M_k\overline{N^c_k}P_R N_k 
+ M_k\overline{N_k}P_L N_k^c\right)
-(h_{\alpha k}\overline{\ell_\alpha}\eta P_R N_k +{\rm h.c.}), \nonumber\\
V&=&m_\phi^2\phi^\dagger\phi+m_\eta^2\eta^\dagger\eta+
\lambda_1(\phi^\dagger\phi)^2+
\lambda_2(\eta^\dagger\eta)^2+
\lambda_3(\phi^\dagger\phi)(\eta^\dagger\eta)+
\lambda_4(\phi^\dagger\eta)(\eta^\dagger\phi) \nonumber \\
&+&
\frac{\lambda_5}{2}\left[(\phi^\dagger\eta)^2+ {\rm h.c.}\right],
\label{nlag}
\end{eqnarray}
where $\ell_\alpha$ and $\phi$ stand for a left-handed lepton doublet and 
an ordinary Higgs doublet scalar, respectively. Coupling constants and
masses of singlet fermions are
assumed to be real, for simplicity.
${\cal L}_N$ is assumed to be written by using the basis in which
a charged lepton mass matrix is diagonal. 

The model has several interesting features \cite{phenom,ks,cms,sty}.
First, neutrino masses are generated through one-loop diagrams as
\begin{eqnarray}
&&{\cal M}_{\alpha\beta}=\sum_k\Lambda_k h_{\alpha k}h_{\beta k},
 \nonumber \\
&&\Lambda_k=\frac{\lambda_5\langle\phi\rangle^2}{8\pi^2M_k}~
I\left(\frac{M_k^2}{M_\eta^2}\right), \qquad
I(x)=\frac{x}{1-x}
\left(1+\frac{x~\ln~x}{1-x}\right),
\label{nmass}
\end{eqnarray}
where $M_\eta^2=m_\eta^2+(\lambda_3+\lambda_4)\langle\phi\rangle^2$.
This neutrino mass matrix can explain the neutrino oscillation data well
as long as we set appropriate values for the parameters 
$\lambda_5$, $h_{\alpha k}$, $M_k$ and $M_\eta$.
We note that $\lambda_5$ should be very small to generate desired neutrino
masses. However, this tuning is not so bad nature since it can be
controlled by a global symmetry which appears if we make $\lambda_5$
zero.\footnote{This problem may also be solved by making the $\lambda_5$ 
term as an effective one through the extension of the model 
with a U(1) gauge symmetry \cite{ks}.}
Second, the lightest one of $N_k$ can be DM since its stability is
guaranteed by the $Z_2$ symmetry.  
Its relic density can be adjusted to the one required by WMAP 
for the same parameters used for the explanation of the neutrino 
oscillation data.
Third, these are also consistent with the constraints imposed by 
the lepton flavor violating processes such as $\mu\rightarrow e\gamma$ 
and $\tau\rightarrow \mu\gamma$, if neutrino Yukawa couplings 
$h_{\alpha k}$ have certain flavor structure.\footnote{There is a severe
tension between the relic abundance and the lepton flavor violation
generally \cite{phenom,ks}. 
If we make neutrino Yukawa couplings small enough to suppress
the lepton flavor violation, the DM relic abundance becomes too large.} 

In order to show these aspects concretely, 
we consider an example of such flavor structure for neutrino Yukawa
couplings as 
\begin{equation}
h_{ei}=0, \quad h_{\mu i}=h_{\tau i}~ (i=1,2); \quad 
h_{e3}=-h_{\tau 3}, \quad \quad h_{\mu 3}=-h_{\tau 3}.
\label{yukawa}
\end{equation}
In this case the neutrino mass matrix can be written as
\begin{equation}
{\cal M}=\left(\begin{array}{ccc}
0 & 0 & 0\\ 0 & 1 & 1 \\ 0 & 1 & 1 \\
\end{array}\right)(h_{\tau 1}^2\Lambda_1+h_{\tau 2}^2\Lambda_2)+
\left(\begin{array}{ccc}
1 & 1 & -1\\ 1 & 1 & -1 \\ -1 & -1 & 1 \\
\end{array}\right)h_{\tau 3}^2\Lambda_3,
\end{equation}
and the tri-bimaximal neutrino mixing is automatically
realized for the neutrino mass matrix (\ref{nmass}) \cite{sty}. 
Moreover, only two mass eigenvalues take nonzero values.
Thus, the neutrino oscillation data can be consistently explained 
as long as the following conditions are satisfied: 
\begin{equation}
h_{\tau 1}^2\Lambda_1+h_{\tau 2}^2\Lambda_2
\simeq 2.5\times 10^{-2}~{\rm eV}, \qquad
h_{\tau 3}^2\Lambda_3\simeq 2.9\times 10^{-3}~{\rm eV}. 
\label{coscil}
\end{equation}
These come from the required values for $\Delta m_{\rm atm}^2$ and
$\Delta m_{\rm solar}^2$, respectively.
We need to consider the constraints from both the lepton flavor violating 
processes and the DM relic abundance under these conditions. 
The relation of Yukawa couplings $h_{\alpha k}$ to other
parameters $\lambda_5,~M_k$ and $M_\eta$ is also determined 
through these constraints. 
When we apply eq.~(\ref{coscil}) to the analysis, it may be useful to note that 
these give the constraints on the value of $h_{\tau k}^2\lambda_5$ for
the fixed $M_k$ and $M_\eta$. In particular,
$\Lambda_k$ is proportional to $h^2_{\tau k}\lambda_5 M_k/M_\eta^2$ 
for $M_k\ll M_\eta$ since $I(x)\simeq x$ for $x\ll 1$.
Since $h_{\tau k}$ tends to be smaller for
larger values of $\lambda_5$, $\lambda_5$ is expected to have values 
in restricted regions by taking account of the DM relic abundance
condition as seen later.  
We will assume $M_1 < M_\eta$ throughout the present analysis. 

\input epsf
\begin{figure}[t]
\begin{center}
\epsfxsize=7cm
\leavevmode
\epsfbox{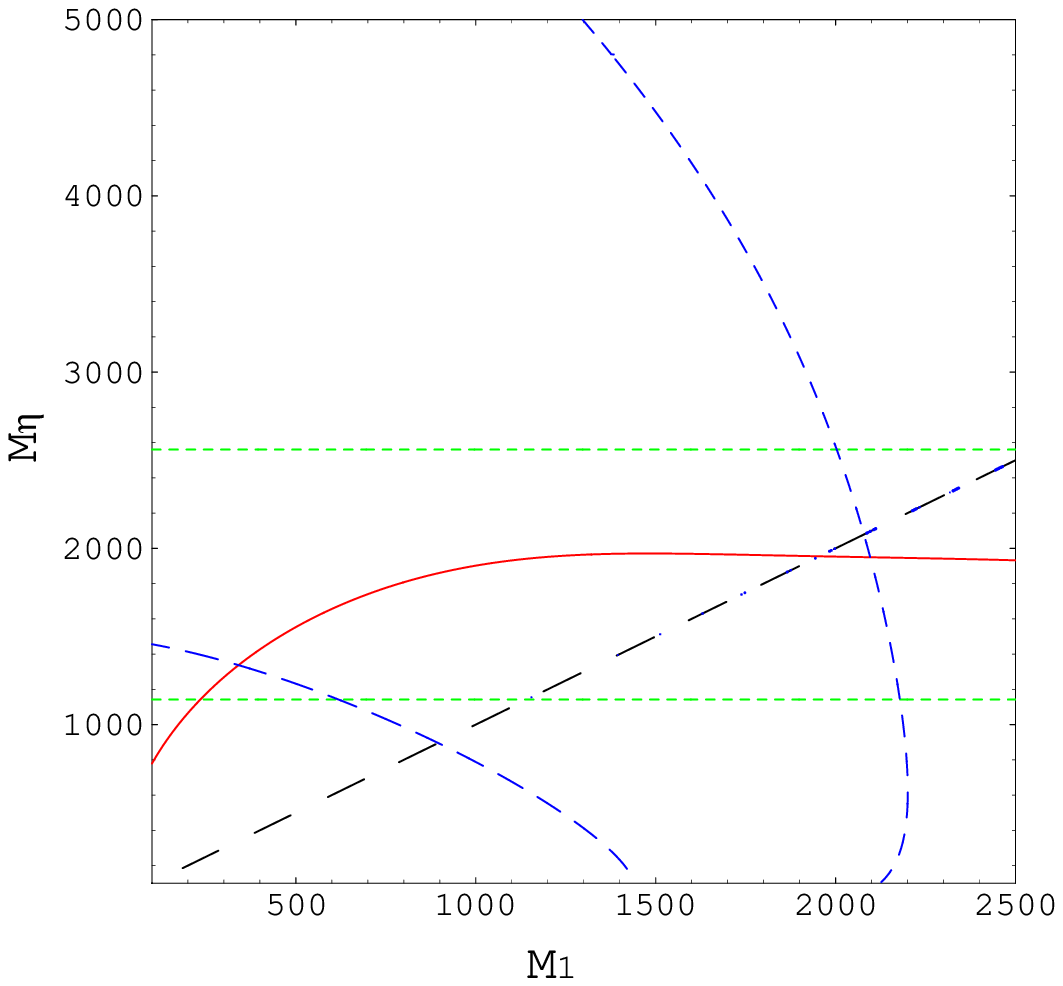}
\hspace*{7mm}
\epsfxsize=7cm
\leavevmode
\epsfbox{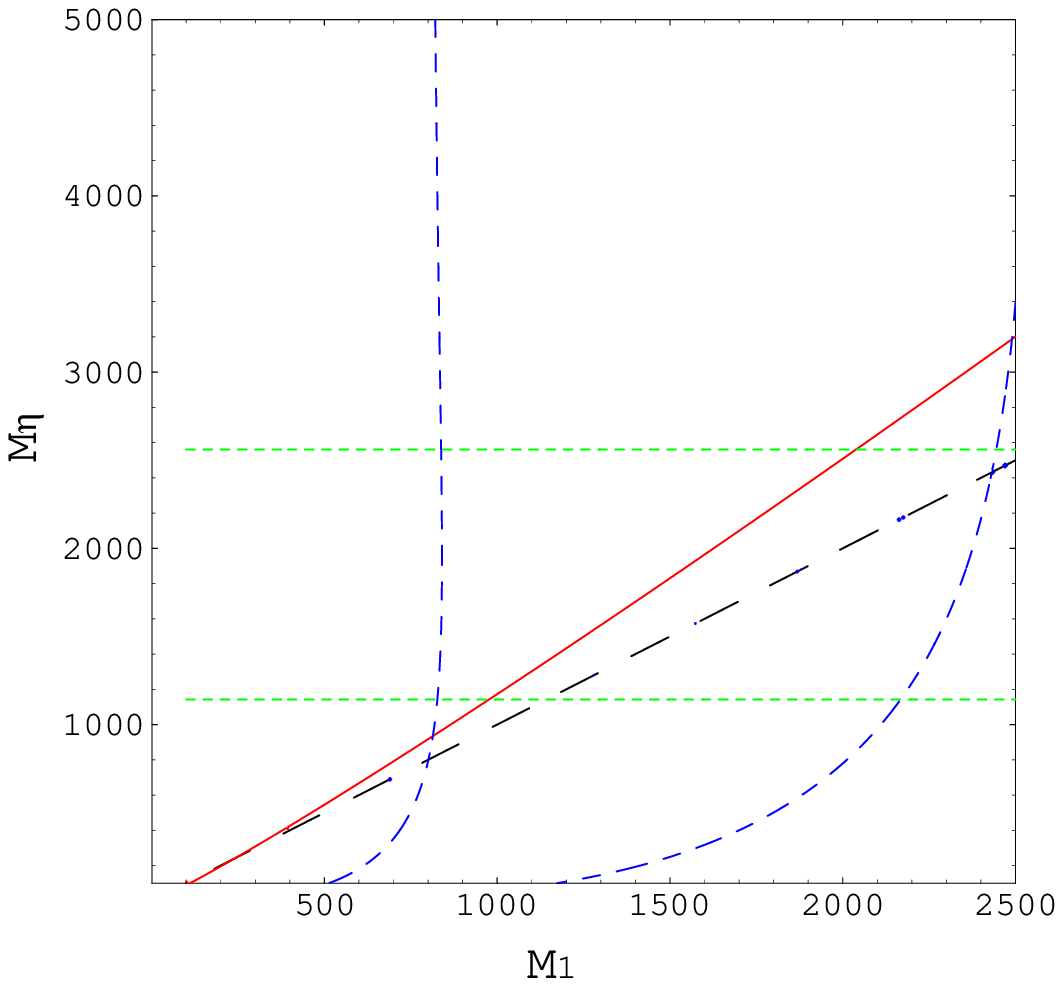}
\end{center}

\vspace*{-3mm}
{\footnotesize {\bf Fig.~1}~~
Contours for the branching ratio of the lepton flavor violating processes
and the DM relic abundance in the $(M_1,~M_\eta)$ plane. 
The left and right panel corresponds to case (i) and (ii) defined in the
 text, respectively. 
Green dotted lines represent the contours for 
$Br(\mu\rightarrow e\gamma)\times 10^{11}=1.2, 0.72$ in $M_\eta$ decreasing
 order. Blue dashed lines represent the ones for 
$Br(\tau\rightarrow \mu\gamma)\times 10^8=0.68, 0.068$ in $M_1$
 increasing order. 
The red solid line in the $M_1<M_\eta$ region corresponds to the contour 
for the $N_1$ relic abundance $\Omega_{N_1}h^2=0.11$ required by the
 WMAP. The black long dashed line represents a line for $M_1=M_\eta$.}
\end{figure}

The branching ratio of the lepton flavor
violating process $\ell_\alpha\rightarrow\ell_\beta\gamma$ is
written as \cite{fmeg}
\begin{eqnarray}
&&Br(\ell^-_\alpha\rightarrow\ell^-_\beta\gamma)
=\frac{3\alpha}{64\pi(G_FM_\eta^2)^2}\left[\sum_{k=1}^3h_{\alpha k}
h_{\beta k}F_2\left(\frac{M_k^2}{M_\eta^2}\right)\right]^2
Br(\ell^-_\alpha\rightarrow\ell^-_\beta\bar\nu_\beta\nu_\alpha),
\nonumber\\
&&F_2(x)=\frac{1-6x+3x^2+2x^3-6x^2\ln x}{6(1-x)^4}.
\end{eqnarray}
If we use the condition (\ref{yukawa}), we find that
\begin{eqnarray}
&&Br(\mu\rightarrow e\gamma)\simeq
\frac{3\alpha}{64\pi(G_FM_\eta^2)^2}\left[h_{\tau 3}^2
F_2\left(\frac{M_3^2}{M_\eta^2}\right)\right]^2, \nonumber \\
&&Br(\tau\rightarrow \mu\gamma)\simeq
\frac{0.51\alpha}{64\pi(G_FM_\eta^2)^2}\left[
h_{\tau 1}^2 F_2\left(\frac{M_1^2}{M_\eta^2}\right)
+h_{\tau 2}^2F_2\left(\frac{M_2^2}{M_\eta^2}\right)
-h_{\tau 3}^2F_2\left(\frac{M_3^2}{M_\eta^2}\right)\right]^2.
\label{eqlfv}
\end{eqnarray}
By using these formulas and eqs.~(\ref{coscil}), the expected values 
for the branching ratio of 
$\mu\rightarrow e\gamma$ and $\tau\rightarrow \mu\gamma$ 
can be plotted in the $(M_1,~M_\eta)$ plane by fixing parameters
$h_{\tau 1},~\lambda_5,~M_2$, and $M_3$.
Here we consider two cases: (i)~$M_1<M_2<M_3$ and 
(ii)~$M_1\simeq M_2<M_3$. In both cases $\lambda_5$ and $M_3$ are
treated as free parameters.  
Since $h_{\tau 1}$ and $M_2$ are determined by other parameters 
in case (ii), this case is much constrained and predictive 
compared with case (i).

In Fig.~1 we show the contours of these branching ratios 
for typical parameters. Here we use $\lambda_5=6.0\times 10^{-11}$ 
and $M_3=4.8$~TeV. In case (i), we fix the remaining parameters as 
$h_{\tau 1}=1.5$ and $M_2=2.8$~TeV.
Green dotted lines and blue dashed lines represent the contours of 
$Br(\mu\rightarrow e\gamma)$ and $Br(\tau\rightarrow\mu\gamma)$, respectively.
The former one is independent of $M_1$ in both cases (i) and (ii).
This is clear from the expression in eq.~(\ref{eqlfv}). Moreover, this
branching ratio becomes sufficiently small by making $M_3$ large enough.
It should be noted that $F_2(M^2_3/M_\eta^2)$ becomes smaller for larger
$M_3$ although larger $M_3$ makes $h_{\tau 3}$ larger 
through eq.(\ref{coscil}). 
On the other hand, $Br(\tau\rightarrow\mu\gamma)$ shows different 
behavior in each case.
It is independent of $M_\eta$ in case (ii) for the $M_\eta>M_1$ region. 
This is expected from the feature of $\Lambda_k$ which is previously remarked 
on eqs.~(\ref{coscil}) and (\ref{eqlfv}).
In case (i), $Br(\tau\rightarrow\mu\gamma)$ is not largely affected
by changing $M_2$ and $M_3$ as long as $h_{\tau 1}>h_{\tau 2}$ is
satisfied, which is favored by the DM relic abundance as seen later.  
The present experimental bounds \cite{clfv} are found to be satisfied 
in the wide range of parameter space shown in this figure.
The model can be easily consistent with both the neutrino oscillation
data and the bounds from the lepton flavor violating processes 
as long as parameters are suitably selected.
  
\begin{figure}[t]
\begin{center}
\epsfxsize=12cm
\leavevmode
\epsfbox{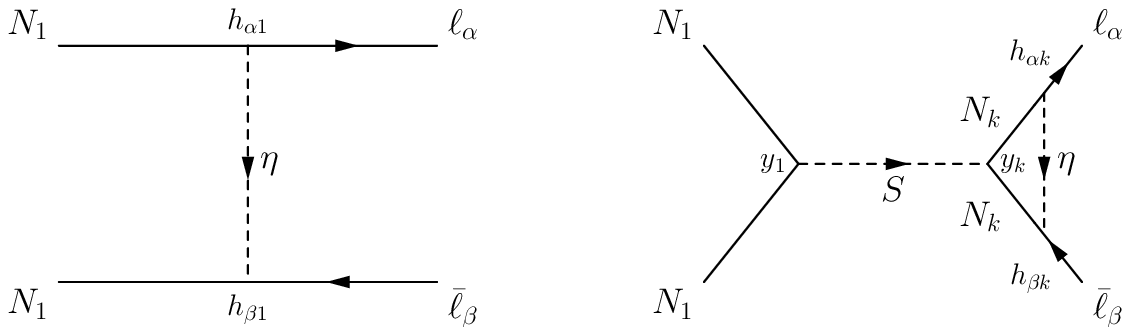}

{\footnotesize {\bf Fig.~2}~~Diagrams contributing to the $N_1$
 annihilation.}
\end{center}
\end{figure}

Next we discuss the nature of DM in this model.
Since $N_1$ is assumed to be DM, the condition (\ref{yukawa}) suggests 
that charged final states yielded in the DM annihilation consist 
of $\mu^\pm$ or $\tau^\pm$ only. 
Positron and electron are only induced through the decay of these particles.
It should be reminded again that this nature of DM is favored by the anomaly
suggested by PAMELA and Fermi-LAT.
The DM relic abundance is determined through the
$N_1$ annihilation, which occurs through the $t$-channel $\eta$
exchange diagram shown in Fig.~2. The dominant contribution comes
from the $p$-wave process.
Thus, the annihilation cross section averaged by the spin of initial 
states is expressed as\footnote{
We need to remind that final states also include neutrino pairs 
other than the charged lepton pairs for the relic abundance estimation.}
\begin{equation}
\sigma_1v=\frac{1}{3\pi}
\frac{M_1^2(M_1^4+M_\eta^4)}{(M_1^2+M_\eta^2)^4}
h_{\tau i_1}^2h_{\tau i_2}^2v^2,
\label{t0cross}
\end{equation}
where we use eq.~(\ref{yukawa}) to derive this formula.
In case (i), $i_1=i_2=1$ should be understood and 
then $\sigma_1v\propto h_{\tau 1}^4$. 
On the other hand, if the masses of $N_1$ and $N_2$ are almost 
degenerate as in case (ii), coannihilation plays a role and 
then the contribution of $i_{1,2}=1$ and 2 should be summed up.
As its result, we have 
$\sigma_1v\propto(h_{\tau 1}^2+h_{\tau 2}^2)^2$ \cite{sty}. 

In order to estimate the freeze-out temperature $T_f$ of $N_1$ including the 
coannihilation case, we follow the procedure given in \cite{dma1,dma2}.
We define $\sigma_{\rm eff}$ and $g_{\rm eff}$ as
\begin{eqnarray}
\sigma_{\rm eff}&=& 
\frac{g_{N_1}^2}{g_{\rm eff}^2}\sigma_{N_1N_1}+
2\frac{g_{N_1}g_{N_2}}{g_{\rm eff}^2}
\sigma_{N_1N_2}(1+\delta)^{3/2}e^{-x\delta}
+\frac{g_{N_2}^2}{g_{\rm eff}^2}\sigma_{N_2N_2}
(1+\delta)^3e^{-2x\delta}, \nonumber \\
g_{\rm eff}&=&g_{N_1}+g_{N_2}(1+\delta)^{3/2}e^{-x\delta},
\label{effcross}
\end{eqnarray}
where $x=M_1/T$ and $m_{\rm pl}=1.22\times 10^{19}$~GeV.
Internal degrees of freedom of $N_i$ are described by $g_{N_i}$ and  
$\delta$ is defined by $\delta\equiv (M_2-M_1)/M_1$. 
If we define $a_{\rm eff}$ and $b_{\rm eff}$ by 
$\sigma_{\rm eff}v=a_{\rm eff}+b_{\rm eff}v^2$, 
the thermally averaged cross section can be written as 
$\langle\sigma_{\rm eff}v\rangle=a_{\rm eff}+ 6 b_{\rm eff}/x$.
In case (i), $\sigma_{\rm eff}$ and $g_{\rm eff}$ are dominated by the
first term since $\delta> 0.2$ \cite{dma2}. On the other hand, $\delta\ll 1$ is
assumed in case (ii). Thus, the second and third terms can bring
the important contribution.
Using these, the relic abundance of $N_1$ can be estimated through 
the formulas
\begin{equation}
\Omega_{N_1} h^2=\frac{1.07\times 10^9 x_f}{g_\ast^{1/2}m_{\rm pl}({\rm
 GeV})
(a_{\rm eff}+ 3b_{\rm eff}/x_f)}, \quad
x_f=\ln\frac{0.038 g_{\rm eff} m_{\rm pl}M_1(a_{\rm eff}
+6b_{\rm eff}/x_f)}{g_\ast^{1/2}x_f^{1/2}},
\end{equation}
where $g_\ast$ is the relativistic degrees of freedom at the freeze-out
temperature $T_f$ of $N_1$. 

By using these formulas and the conditions in eq.~(\ref{coscil}), 
we can plot the contour $\Omega_{N_1}h^2=0.11$ required by WMAP 
in the $(M_1, M_\eta)$ plane. 
In Fig.~1, it is drawn by a red solid line in each case (i) and (ii)
for the same values of parameters used in the estimation of 
$Br(\ell_\alpha\rightarrow\ell_\beta\gamma)$.
The result in each case depends on $h_{\tau 1}^2$ 
and $h_{\tau 1}^2 +h_{\tau 2}^2$, respectively. 
Since it also depends on $\lambda_5$ through the relations
(\ref{coscil}) as noted before, the required $\Omega_{N_1} h^2$
can be obtained only for rather restricted values of $\lambda_5$.
In Fig.~1, the points on the red solid line in the region satisfying 
both $Br(\mu\rightarrow e\gamma)<1.2\times 10^{-11}$ and $M_1<M_\eta$ 
give the parameters consistent with 
all of the neutrino oscillation data and constraints from the lepton 
flavor violating process and the DM relic abundance.
Thus, we find that the present model can give a very simple 
and consistent framework 
for the known experimental results.

The values of relevant Yukawa couplings are determined for each 
point on the $\Omega_{N_1}h^2=0.11$ line. In Fig.~1, we have, for example,
\begin{equation}
\begin{array}{lll}
{\rm (i)}& h_{\tau 2}=1.40, \quad h_{\tau 3}=0.66 &
 \quad  {\rm at}~ (M_1,M_\eta)=(1600,1950), \\
{\rm (ii)}&\sqrt{h_{\tau 1}^2+h_{\tau2}^2}=2.14, \quad h_{\tau 3}=0.66 &
 \quad  {\rm at}~ (M_1,M_\eta)=(1600,1950). \\
\end{array}
\label{yukawa1}
\end{equation}
If we make $\lambda_5$ larger for the fixed $M_1$, larger values for 
$M_\eta$ and $h_{\tau 1}$ are required as expected from eq.~(\ref{t0cross}). 
On the other hand, $\lambda_5$ is bounded from below by the condition
$M_\eta>M_1$. 
Thus, neutrino Yukawa couplings are required to take rather large
values by the DM relic abundance. 
This suggests that the model may be inconsistent 
due to these large Yukawa couplings, which may make the scalar potential 
unstable at the energy regions above a certain cut off scale $\mu$.  
If this cut-off scale does not satisfy $\mu>M_3$, the present scenario 
can not work. Since larger $M_3$ is favored from the 
$\mu\rightarrow e\gamma$ constraint, we can not make $M_3$ smaller
enough for this instability problem. Thus, this imposes nontrivial
constraint on the model. 
Since $\lambda_2$ is most affected by the large neutrino
Yukawa couplings $h_{\tau k}$, 
the cut-off scale $\mu$ is determined as the scale where $\lambda_2$ 
becomes negative. 

We examine this point by studying the behavior of couplings included in 
${\cal L}_N$ and $V$ to fix $\mu$ using renormalization group 
equations (RGEs) for them. These RGEs are given in Appendix A.
Numerical analysis is practiced for the parameters given 
in (\ref{yukawa1}) assuming the $O(1)$ values for the couplings $\lambda_i$.
This analysis shows that $\mu=O(10)$~TeV and then 
$\mu>M_3$ is possible for these values of couplings 
at low energy regions. The scenario seems to be consistent with 
the potential instability.
However, it is difficult to make $\mu$ much larger than $M_3$.
If $N_k$ and $\eta$ are supposed to be suitable representations 
of some hidden non-Abelian gauge symmetry under which all the SM contents 
are singlet, some improvement may be expected for this situation. 
As such an example, we may consider SU(2) symmetry
and both $N_k$ and $\eta$ are doublet of that gauge symmetry. 
In such an extension, ${\cal L}_N$
and $V$ are invariant and no anomaly problem occurs within these field
contents. The RGE study of this case shows that $\mu$ can be somewhat 
large.
However, it is difficult to make $\mu $ larger than $M_3$ by more than
one order since the running region of the relevant RGEs is too short.
Thus, although the model can escape the instability of the potential, 
we need to consider some fundamental model at the scale not far from $M_3$.
Since this argument on the potential stability suggests that smaller 
neutrino Yukawa couplings are favored, smaller values of $M_1$ 
and $\lambda_5$ are also favored from the $N_1$ relic abundance.
On the other hand, as discussed in the next part,
only a limited value of $M_1$ seems to be favored from the explanation 
of the charged cosmic ray anomaly.
This suggests that $\lambda_5$ is also required to take its value in the
strictly restricted region.  

It is worthy to note that we can predict the expected values 
for the branching ratio of $\mu\rightarrow e\gamma$ and 
$\tau\rightarrow\mu\gamma$ in this model from Fig.~1, 
if we can fix the value of $M_1$ further by using other experimental data.
Observational data on the cosmic rays from PAMELA and Fermi-LAT experiments 
may be used for such a purpose. 
However, if we suppose the PAMELA anomaly as a consequence of 
the annihilation of this DM, we confront difficulty.
The annihilation cross section is found to be too small 
to explain the PAMELA positron excess for the typical relative 
velocity of DM in the present Galaxy as mentioned before.
In the next part we propose an extension 
of the model to overcome this fault.

\subsection{Extension of the model}
We consider the introduction of a new interaction which brings a large 
contribution to the $N_1$ annihilation only at the present Galaxy 
and also does not modify the previously discussed favorable features 
of the model. For that purpose, we add a complex singlet scalar $S$ with even
parity of the $Z_2$ symmetry.\footnote{The extension of the model by a
singlet scalar field has been considered in other context in \cite{bm}.}
 This singlet scalar is assumed to have 
mass $M_S$ and couplings with other fields through the Lagrangian 
through the interaction Lagrangian
\begin{equation}
{\cal L}_N^\prime= -y_kS\overline{N^c_k}P_R N_k 
-y_k^\ast S^\ast\overline{N_k}P_L N_k^c
-M_S^2|S|^2-\kappa|S|^4-
\left(\kappa_\phi\phi^\dagger\phi+\kappa_\eta\eta^\dagger\eta\right)|S|^2.
\label{addlag}
\end{equation}
Here we note that this is not the most general Lagrangian
under the imposed symmetry.
However, although interaction terms like $\phi^\dagger\phi S$ and
$\eta^\dagger\eta S$ which are not forbidden by the symmetry can be
radiatively induced, they are largely suppressed as long as $S$ is
assumed to have no vacuum expectation value.\footnote{This assumption 
is justified only if
the tadpole diagram for $S$ generated through the $N_k$ loop is
cancelled by $cS$ which can be introduced in Lagrangian. We consider
such a situation here.} 
In that case, $S$ dominantly decays to $N_k$ with the mass $M_k<M_S/2$.
In this extended model, we find that there appears a new one-loop
contribution to the annihilation of $N_1$ as shown in Fig.~2. 

This new contribution to the $N_1$ annihilation cross 
section can be estimated as
\begin{equation}
(\sigma_2 v)_{\alpha\beta}=\frac{1}{\pi}
\frac{M_1^2}{M_\eta^4}
\frac{m_\alpha^2+m_\beta^2}{(s-M_S^2)^2+M_S^2\Gamma_S^2}
\left(\sum_{k=1}^3\frac{|y_1y_k|h_{\alpha k}h_{\beta k}M_k}
{(4\pi)^2 D_k}\right)^2,
\label{one-loop}
\end{equation}
where the spin of initial states is averaged.  
We fix the final states to be charged leptons with masses 
$m_{\alpha,\beta}$ in this expression.
This annihilation cross section is dominated by the contribution from
 the exchange of a pseudoscalar component. 
To obtain the total annihilation cross section, $\alpha$ and $\beta$
should be summed up for all possible final states as
$\sigma_2v=\sum_{\alpha\beta}(\sigma_2v)_{\alpha\beta}$.  
The definition of $s$, $\Gamma_S$ and $1/D_k$ are given by
\begin{eqnarray}
&& s=E_{\mathrm{cm}}^2\simeq 4M_1^2\left(1+\frac{v^2}{4}\right),\nonumber\\
&&\Gamma_S=\frac{|y_1|^2}{8\pi}M_S
\sqrt{1-4\frac{M_1^2}{M_S^2}}\left(1-2\frac{M_1^2}{M_S^2}\right), \nonumber \\
&&\frac{1}{D_k}=\int^1_0dz~\frac{1}{1-r_k+4r_1z}
\log{\left|\frac{1-(1-r_k)z}
{r_k-4r_1z(1-z)}\right|},
\end{eqnarray} 
where $r_k=M_k^2/M_\eta^2$ and $S$ is supposed to decay to the $N_1$ pair
only. 
This type of annihilation cross section has been suggested to be
enhanced sufficiently for the explanation of the PAMELA data \cite{bw}.
In fact, if the thermal average of $(\sigma v)_{\alpha\beta}$ is estimated
naively by replacing $v^2$ with a thermally averaged value $\frac{6}{x}$ 
in eq.~(\ref{one-loop}), the annihilation cross section shows 
the Breit-Wigner resonance at $x_r=\frac{3}{2\Delta}$ through the factor 
$[(\Delta -\frac{3}{2x})^2+\gamma_S^2]^{-1}$, where we use the definition 
$\Delta\equiv 1-\frac{4M_1^2}{M_S^2}$ and 
$\gamma_S\simeq\frac{1}{16\pi}|y_1|^2\Delta^{1/2}$. 
However, such a naive treatment has been shown to be unreliable near a
resonance point \cite{dma2,tav}. 
The enhancement of the annihilation cross section is overestimated 
in such a naive method.
To obtain the correct enhancement, we need to calculate the thermal average
\begin{equation}
\langle\sigma_2v\rangle_{\alpha\beta}=\frac{x^{3/2}}{2\pi^{1/2}}
\int^\infty_0 dv~v^2(\sigma_2v)_{\alpha\beta}e^{-xv^2/4}.
\label{num}
\end{equation}
Although this formula is derived in the center of mass system, the result is
expected to be reliable since $N_1$ is sufficiently non-relativistic in
the present case \cite{dma2,tav}.

In order to find the qualitative feature, it is useful to approximate 
this integral by expanding $v$ as $v=v_r+\nu$ around the peak value 
$v_r=2\Delta^{1/2}$. Then, eqs.~(\ref{one-loop}) and (\ref{num}) give 
\begin{eqnarray}
\langle\sigma_2v\rangle_{\alpha\beta}&\simeq&\frac{x^{3/2}}{2\pi^{3/2}}
\frac{(m_\alpha^2+m_\beta^2)M_1^2}{M_\eta^4M_S^4}
\left(\sum_{k=1}^3\frac{|y_1y_k|h_{\alpha k}h_{\beta k}M_k}
{(4\pi)^2 D_k}\right)^2 e^{-x\Delta}\int^{\nu_0}_{-\nu_0}d\nu
\frac{1}{\nu^2\Delta+\gamma^2_S} \nonumber \\
&\simeq&\frac{2\pi^{1/2}}{(4\pi)^4}\frac{(m_\alpha^2+m_\beta^2)}{M_\eta^4}
\left(\sum_{k=1}^3\frac{|y_k|h_{\alpha k}h_{\beta k}M_k}
{D_kM_1}\right)^2 x^{3/2}e^{-x\Delta},
\label{cross}
\end{eqnarray}
where $\nu_0\ll v_r$ and $16\pi\nu_0\gg |y_1|^2$ are assumed.  
Using this result, we roughly estimate this resonance effect 
on the annihilation cross section caused by the diagram which 
has $N_k$ as the internal fermions and $\tau^\pm$ in final states.
For that purpose, we take $x_r\simeq 10^6$ which is just 
coincident with the typical relative velocity $2\times 10^{-3}c$ of 
this DM in the present Galaxy. 
The annihilation cross section at $x_r$ is found to satisfy the relation 
\footnote{Here the annihilation cross section
$\langle\sigma_2v\rangle$ is defined as 
$\langle\sigma_2v\rangle=4\langle\sigma_2v\rangle_{\mu^\pm\tau^\mp}$
by taking account of all possible modes. See eqs.(\ref{pflux2}) 
and (\ref{tcross}) also.}
\begin{equation}
\frac{\langle\sigma_2v\rangle}{10^6\langle\sigma_1v\rangle}
\sim \left(\frac{M_k}{M_1D_k}\right)^2
\left(\frac{h_{\tau k}}{\tilde h}\right)^4 |y_k|^2,
\label{boost}
\end{equation}
where $\tilde h=h_{\tau 1}$ and $\sqrt{h_{\tau 1}^2+h_{\tau 2}^2}$ for
the annihilation and the coannihilation, respectively.
The first two factors relevant to the masses and the neutrino Yukawa couplings 
of $N_k$ are fixed by the conditions imposed by the neutrino oscillation
and the $N_1$ relic abundance. 
The first factor is estimated to be $O(1)$ and decreases for larger $M_k$.
The second factor is considered to be less than 1
except for the coannihilation case where it can be almost 1. 
Since $|y_1|$ is assumed small in the above discussion, we find that 
the desirable enhancement can be expected from the $N_2$ contribution with 
$|y_2|=O(1)$.
These show that the sufficient enhancement factor to explain the PAMELA
data can be obtained through the Breit-Wigner resonance at
least in the coannihilation case.

\begin{figure}[t]
\begin{center}
\epsfxsize=7cm
\leavevmode
\epsfbox{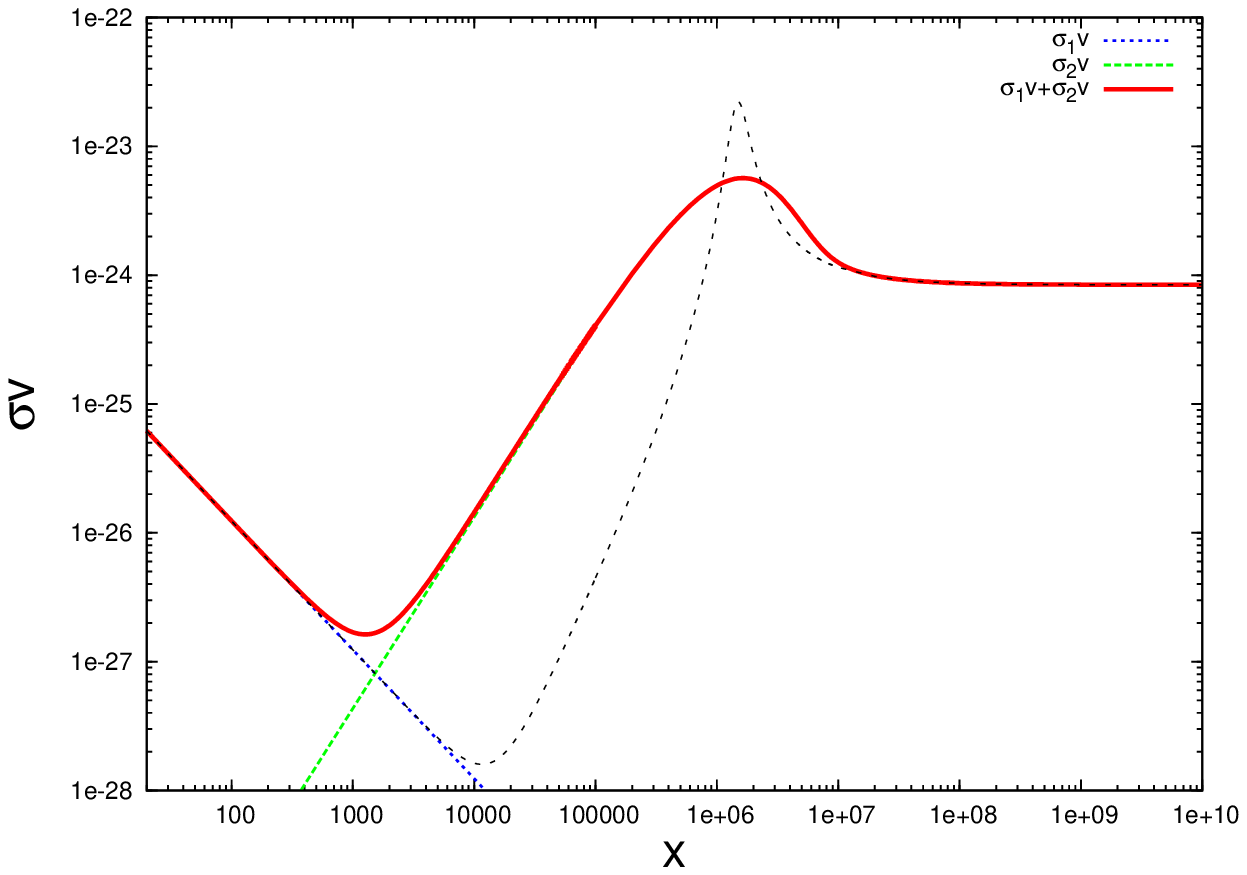}
\hspace*{5mm}
\epsfxsize=7cm
\leavevmode
\epsfbox{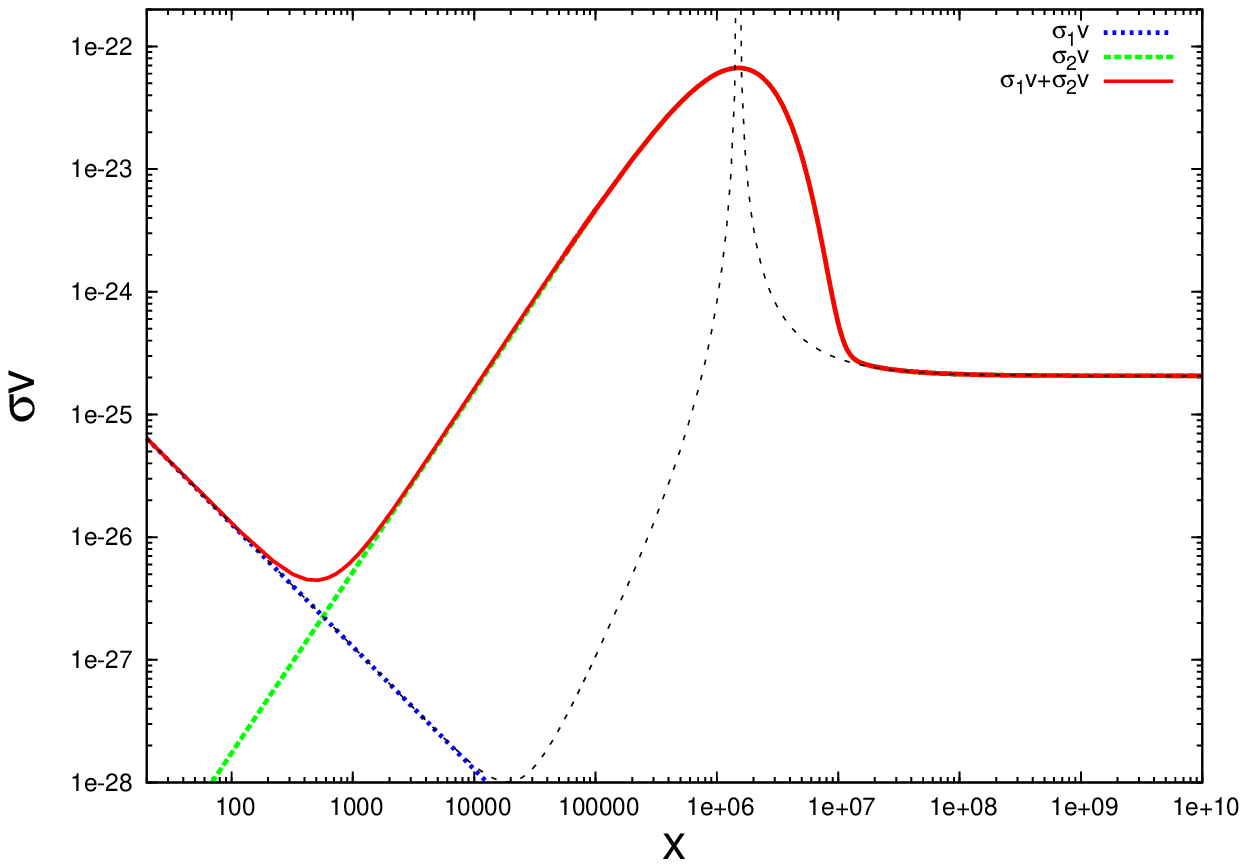}
\end{center}

{\footnotesize {\bf Fig.~3}~~The $N_1$ annihilation cross section as
 a function of $x(\equiv\frac{6}{\langle v^2\rangle})$. 
The left and right panel
 corresponds to the case (i) and (ii), respectively.
Parameters in the annihilation cross section are fixed to the ones 
shown in Table~1. The thin black dashed line shows the result for 
$\langle\sigma v\rangle$ obtained by the naive method.}
\end{figure}

\begin{figure}[b]
\begin{center}
\begin{tabular}{cccccccccc}\hline
&$M_1$&$M_\eta$ & $h_{\tau 1}$ &$h_{\tau 2}$ &$h_{\tau 3}$ & $\Delta$ &
 $|y_1|$ & $|y_2|$ & $|y_3|$\\\hline  
(i) & 1.6& 1.95& 1.5 & 1.4  & 0.66  & $10^{-6}$  & 0.1 & 2.5
				 & 0.01\\
(ii)& 1.6&1.95& 0.1  & 2.14  & 0.66  & $10^{-6}$  & 0.015 &
				 1.715 & 0.01 \\\hline
\end{tabular}
\end{center}
\vspace*{4mm}

\footnotesize{{\bf Table~1}~~ Parameter sets used to draw the
 annihilation cross section behavior in Fig.~2 and also 
to obtain the positron spectrum in Fig.~4. Masses are given in TeV unit.
We set $\lambda_5=6.0\times 10^{-11}$ and $M_3=4.8$~TeV for each case.}
\vspace*{8mm}
\end{figure}

To obtain much quantitative estimation we calculate the thermally 
averaged annihilation cross section 
by integrating eq.~(\ref{num}) numerically. The result for 
$\langle\sigma v\rangle=\langle\sigma_1 v\rangle+\langle\sigma_2 v\rangle$ 
is plotted as a function of $x$ in Fig.~3.
In this calculation we use the parameters given in Table 1,
which can realize a point on the red line in Fig.~1. 
They satisfy all the neutrino oscillation data, 
the DM relic abundance required by WMAP and the constraints from 
the lepton flavor violating processes.
Since the interference terms between tree diagrams and one-loop
diagrams can be neglected in both regions $v_f/c\sim 0.2$ and
$v_r/c~{^<_\sim}~ 10^{-3}$, the figure shows that we can safely use 
$\langle\sigma_1v\rangle$ and $\langle\sigma_2v\rangle$ in each region, 
respectively. 
In this figure the result obtained by the naive method is also plotted by a
thin black dashed line. It shows that the enhancement effect is
overestimated and the annihilation cross section is misled to be 
large enough for the explanation of the PAMELA data 
in both cases (i) and (ii). 
However, the correct calculation shows that the enhancement can not
be large enough for reasonable values of $|y_k|$ in the annihilation case (i).
On the other hand, in the coannihilation case (ii), 
we find that the Breit-Wigner resonance can make the annihilation cross 
section have a desirable value $10^{-23}$~cm$^3$/sec around $v\simeq v_r$ 
as long as $\Delta$ and $|y_k|$ have suitable values.
Here we should note that this value of $\Delta$ requires $M_S$ to be
finely tuned to $M_1$ at the level of $O(10^{-6})$.
We also comment on the instability of this solution induced by radiative
corrections. A dominant correction to the singlet scalar mass $M_S$ at the
one-loop level is roughly estimated as $\delta M_S^2\simeq 
\frac{y_2^2}{(4\pi)^2}\mu^2$ where $\mu$ is the cut-off scale of the model.
Since $\mu$ is rather small and $\mu=O(10)$~TeV as discussed before, 
we find that $\delta M_S^2$ is the same order value as the required $M_S$. 
This means that we need the fine tuning of $O(10^{-6})$ to keep 
the stability of this solution from the radiative corrections.
The large value of $y_2$ may require the fine tuning up to the
eight-loop order. 
Unfortunately, the model does not have any physical background 
to guarantee the required mass relation. It remains as a difficult problem 
how to realize this finely tuned situation from the basic model at 
high energy regions.

It is worthwhile to stress that these values of $\Delta$ and $|y_k|$ 
can be fixed without contradicting 
the required DM relic abundance which is determined by the annihilation
process described by eq.~(\ref{t0cross}). 
The reason is that different parameters are relevant to determine
the DM relic abundance and the positron flux, respectively.\footnote{
It is useful to note that the similar aspect is found in the case 
of Sommerfeld enhancements. If only a single annihilation channel 
is assumed, Sommerfeld enhancements cause a discrepancy between the
relic density and the excesses of positron flux \cite{som}.
The present model escapes this by considering two processes given in Fig.~2.}
Although the former is determined by $M_1$, $M_\eta$ and 
$h_{\tau k}$, the latter is mainly determined by $|y_k|$ and $\Delta$.
We note that the parameters relevant to 
the enhancement of the annihilation cross section required for 
the explanation of PAMELA and FERMI-LAT are confined to 
$y_1$, $y_2$ and $M_S$, although a lot of free parameters seems 
to be introduced in eq.~(\ref{addlag}).  
It seems to be interesting that these limited parameters can also allow the
model to satisfy the reionization constraints as discussed 
below.\footnote{The relevant parameters contained in eq.~(\ref{nlag}) have
already been fixed to explain the neutrino oscillation data 
(two squared mass differences and three mixing angle), 
the DM relic abundance $\Omega h^2$ and lepton flavor violating
processes. Taking account of the supposed flavor structure, they are 
$M_{1,3}, M_\eta, \lambda_5$ and $h_{\tau_1}$,
$h_{\tau_2}, h_{\tau_3}$ in the coannihilation case.}

It is also useful to note that it is crucial that the annihilation occurs
through a one-loop diagram in the present enhancement mechanism. 
This is clear from the fact that the enhancement is caused by 
the existence of $N_k$, which satisfies 
$|y_k|\gg|y_1|$ and $h_{\tau k}\sim \tilde h$.\footnote{It is worthy to
note that numerical calculation shows that $\langle\sigma_2v\rangle$ 
has the largest value
for $y_1\simeq 0.1$ and 0.01 at $x=x_r$ in the case (i) and (ii)
respectively. It slowly decreases for larger $|y_1|$.} 
Although these conditions can be satisfied in the coannihilation case, 
in the annihilation case larger $h_{\tau 2}$ requires 
relatively smaller $h_{\tau 1}$ as seen from eq.(\ref{coscil}).
However, small $h_{\tau 1}$ contradicts the condition imposed 
by the $N_1$ relic abundance.  
Thus, only the coannihilation case can realize
$\langle\sigma_2v\rangle\sim 10^{-23}$~cm$^3$/sec 
for each $M_1$ by adjusting the values of $|y_k|$ and $|y_1|$ without 
affecting the $N_1$ relic abundance. 

One may worry about the potential instability caused by the large value
of $|y_2|$ shown in Table 1. In fact, the coupling constants $\kappa$
and $\kappa_\eta$ can become negative at a scale smaller than $M_3$ as long
as $\eta$ and $N_k$ are singlets of the hidden gauge symmetry.
However, if they are doublets of the hidden SU(2), the coupling constant
$y_k$ in $\langle\sigma_2v\rangle$ is replaced by 
$2y_k$ because of the gauge freedom in the one-loop diagram.
This shows that a rather small value $y_2\simeq 0.86$ is needed to
realize the required enhancement of $\langle\sigma_2v\rangle$. 
We can numerically check that this value of $y_2$ improves the above 
mentioned potential instability problem to make the extended model consistent.
In this case the cut-off scale of the model is still determined 
by the behavior of $\lambda_2$. 

Finally we note the values of the 
annihilation cross section at the recombination period $z\sim 1000$,
which corresponds to the DM relative velocity 
$v/c\sim 10^{-8}$. The DM annihilation in the period after recombination
to structure formation ($z~{^>_\sim}~6$) causes the deposition of 
energy in the inter galactic medium, 
which brings an additional origin for the reionization and heating of the
intergalactic gas. 
This additional effect is constrained from the observed optical depth of
the universe and the measured temperature of the intergalactic gas.
In particular, the optical depth bound brings severe constraint on the
high mass DM as the present model since it can produce too many free electrons. 
If we follow the analysis for these constraints given in \cite{gamma3}, 
the annihilation cross section 
should satisfy $\langle\sigma_2v\rangle~{^<_\sim}~10^{-24}$~cm$^3$/sec for the
DM with the mass $1600$~GeV. In Fig.~3, we find that this constraint 
is satisfied at $v/c~{^<_\sim}~ 10^{-5}$. It 
corresponds to the environments in which most of the annihilation
contribution to the relevant signal is considered to take place. 
Here it is useful to note that $\langle\sigma_2v\rangle$ does not
decrease to $10^{-24}$~cm$^3$/sec even for much smaller relative 
velocity $v<v_r$ and keep larger values than that if
$|y_1|~{^<_\sim}~0.05$ 
is not satisfied.
Thus, the reionization constraint rules out these cases. 
As long as this condition is satisfied, the present DM scenario can 
be consistent with the constraint caused by the effect on the 
reionization due to their annihilation.\footnote{Since the Sommerfeld
enhancement shows an inverse proportionality to the relative velocity of
the two DM fields, it could cause different effects on the reionization
from this model. } 
In the next section we apply this extended model to the explanation of 
the anomaly suggested in PAMELA and Fermi-LAT experiments.

\section{Positron flux and gamma ray constraints}

We estimate the positron flux yielded by the $N_1$ annihilation 
following the method used in \cite{dif,para} and 
compare it with the data obtained in the PAMELA and Fermi-LAT experiments.
The positron flux in the cosmic ray at the Earth is expressed as 
$\Phi_{e^+}(E)=v_{e^+}f(E)/4\pi$~$({\rm GeV}\cdot{\rm cm}^2\cdot{\rm
str}\cdot{\rm sec})^{-1}$ where $v_{e^+}$ is  positron velocity.
$f(E)$ is the positron number density per unit energy at the Earth, which
can be determined by solving the diffusion equation for $f(E)$. 
Using the approximated solution for $f(E)$, the positron flux
$\Phi_{e^+}$ expected from the $N_1$ annihilation is estimated as
 \begin{equation}
\Phi_{e^+}(E)=\frac{v_{e^+}}{8\pi E^2/({\rm GeV}~\tau_E)}\left(\frac{\rho_{N_1}}
{M_1}\right)^2
\int^{M_1}_EdE^\prime ~I(\lambda_D(E,E^\prime))
\left\{ \sum_{\cal F}\langle\sigma v\rangle_{\cal F}
\frac{dN_{\alpha({\cal F}),e^+}}{dE^\prime}\right\},
\label{pflux}
\end{equation}
where $\tau_E=10^{16}$~sec and $\rho_{N_1}$ is the 
local DM density in the halo. In this study we use
$\rho_{N_1}=0.3$~GeV/${\rm cm}^3$ and $v_{e^+}=c$.
Possible final states directly yielded through the $N_1$ 
annihilation are expressed by ${\cal F}$.
$dN_{\alpha({\cal F}),e^+}/dE^\prime$ represents 
the spectrum of positrons yielded through the decay of 
leptons $\alpha$ included in the final state ${\cal F}$. 
 
In this formula, the ingredients coming from astrophysics are summarized in
the halo function $I(\lambda_D)$ and the positron diffusion length 
$\lambda_D$. They are defined by 
\begin{eqnarray}
&&I(\lambda_D)=a_0+a_1\tanh\left(\frac{b_1-\ell}{c_1}\right)
\left\{a_2\exp\left(-\frac{(\ell-b_2)^2}{c_2}\right)+a_3\right\},
\nonumber \\
&&\lambda_D^2=4K_0\tau_E\left\{\frac{E^{\delta-1}-E^{\prime(\delta-1)}}
{1-\delta}\right\},
\label{difm}
\end{eqnarray}
where $\ell=\log_{10}(\lambda_D/{\rm kpc})$. The expressions of
$I(\lambda_D)$ and $\lambda_D$ depend on the astrophysical model for
the diffusion of positron and the halo profile \cite{para}. 
In this paper we adopt med and isothermal profile
for them to determine the parameters included in eq.~(\ref{difm}).
For such a model \cite{para}, parameters in $\lambda_D$ are 
$K_0=0.0112$~kpc$^2$/Myr and $\delta=0.70$, and others included in 
$I(\lambda_D)$ are   
\begin{eqnarray}
&&a_0=0.495,\quad a_1=0.629, \quad a_2=0.137,\quad a_3=0.784, \nonumber \\
&&b_1=0.766,\quad b_2=0.550, \quad c_1=0.193,\quad c_2=0.296. 
\end{eqnarray} 
As addressed in several work \cite{mindep}, the positron flux is not crucially
dependent on the astrophysical model. We choose this model for the
consistency with the constraint from the diffuse gamma in the cosmic ray. 
We will come back to this point later.

In eq.~(\ref{pflux}) the dependence on the assumed model for 
particle physics is confined in the factor summed up for ${\cal F}$ 
in the $E^\prime$ integral. Since the annihilation cross section
$\langle\sigma_2v\rangle_{\alpha\beta}$ is proportional to 
$m_{\alpha}^2+m_\beta^2$, the summation should be taken for 
\begin{equation}
{\cal F}=(e^\pm,\tau^\mp), ~(\mu^\pm,\tau^\mp), ~(\tau^+,\tau^-),
\label{interm}
\end{equation}
which can yield positrons finally.
This feature is caused by the flavor structure of neutrino Yukawa 
couplings (\ref{yukawa}). Since smaller $h_{\tau 3}$ and larger $M_3$ are
favored from the $\mu\rightarrow e\gamma$ constraint, the $N_3$ contribution
to the loop effect may be neglected. 
If we take account of these and also assume that $|y_3|$ is sufficiently
small\footnote{Under this assumption, we can safely neglect the $N_3$
contribution to the one-loop annihilation diagram. In this case $N_1$
annihilation does not yield positrons directly.},
the positron flux $\Phi_{e^+}$ due to the $N_1$ annihilation
can be expressed as
\begin{eqnarray}
\Phi_{e^+}&\simeq& 1.25\times 10^{-3}
\langle\sigma_2v\rangle\left(\frac{10^2~{\rm GeV}}{E}\right)^2 
\left(\frac{1~{\rm TeV}}{M_1}\right)^2 \nonumber \\
&\times&\int_E^{M_1}dE^\prime~I(E,~E^\prime)
\left[\frac{1}{4}\frac{dN_{\mu^+,e^+}}{dE^\prime}
+\frac{3}{4}\left(\frac{dN_{\tau^+,e^+}}{dE^\prime}
+\frac{dN_{\tau^-,e^+}}{dE^\prime}
\right) \right],
\label{pflux2}
\end{eqnarray}
where $({\rm GeV}\cdot{\rm cm}^2\cdot {\rm str}\cdot{\rm sec})^{-1}$
is used for the unit of $\Phi_{e^+}$
and the total cross section $\langle\sigma_2v\rangle
(\equiv 4\langle\sigma_{\mu^\pm\tau^\mp}\rangle)$ is 
determined in our extended model as
\begin{equation}
\langle\sigma_2v\rangle=\frac{16}{(4\pi)^5}\frac{{\rm (GeV)}^2}
{\left(\Delta -\frac{v_r^2}{4}\right)^2+
\gamma_S^2}\frac{M_1^2m_\tau^2}{M_\eta^4M_S^4}
\left(\sum_{k=1}^2\frac{y_1y_kh_{\tau k}^2M_k}{D_k}\right)^2.
\label{tcross}
\end{equation}
Here it should be noted that we can keep the favorable feature 
such that the final states of the $N_1$ annihilation consist of 
heavier leptons only.\footnote{Although higher order radiative 
corrections can induce coupling of $S$ with the ordinary Higgs scalars,
their effect is small enough to neglect them in the analysis 
of $N_1$ annihilation.}  
The fact that $e^\pm$ are not directly produced is favored to
explain the Fermi-LAT data, which show no bump in the hard $e^++e^-$ spectrum.  
The directly produced $e^\pm$ tend to be much harder than indirectly 
produced $e^\pm$ energetically. The positron spectrum given by 
(\ref{pflux2}) has large contributions from $\tau^+$ decay, which causes a 
softer spectrum for the final positron and electron spectrum.
The decay of $\mu^\pm$ to $e^\pm$ yields much harder positron
than the $\tau^\pm$ decay. 
The concrete model with these mixed final states 
seems not to have been considered 
in the analysis of the anomaly suggested through the PAMELA and 
Fermi-LAT experiments.\footnote{Although model independent analysis 
for this kind of mixed final states is found in the paper by Meade {\it et al.} 
\cite{mindep} (see Fig.~12 in it), it is not based on a concrete
particle physics model.}

The energy spectrum of positron $\displaystyle
\frac{dN_{\alpha,e^+}}{dE}$ can be computed by using the PYTHIA 
Monte Carlo code \cite{code}.  
We determine the positron spectrum by fitting these simulation data 
for both the $\mu^+$ and $\tau^\pm$ cases. Details of the analysis 
are given in Appendix B.
We apply this result to eq.~(\ref{pflux2}) to find the positron flux
$\Phi_{e^+}$.
We fix parameters included in the cross section
$\langle\sigma_2v\rangle$ by using the ones 
which realize the point in the allowed region shown in Fig.~1.  
They are also summarized in Table~1 for the case of $M_1=1.6$~TeV.
As expected background fluxes for positrons and electrons, 
we use the empirical formulas given in \cite{bkg},
\begin{eqnarray}
&&\Phi^{\rm bkg}_{e^+}=N_\Phi\frac{4.5E^{0.7}}{1+650E^{2.3}+1500E^{4.2}}, 
\nonumber \\
&&\Phi^{\rm bkg}_{e^-}=N_\Phi\frac{0.16E^{-1.1}}{1+11E^{0.9}+3.2E^{2.15}}
+N_\Phi\frac{0.70E^{0.7}}{1+110E^{1.5}+600E^{2.9}+580E^{4.2}},
\end{eqnarray}
where $E$ should be understood in a unit of GeV and $N_\Phi$ is a
normalization factor.

\begin{figure}[t]
\begin{center}
\epsfxsize=7cm
\leavevmode
\epsfbox{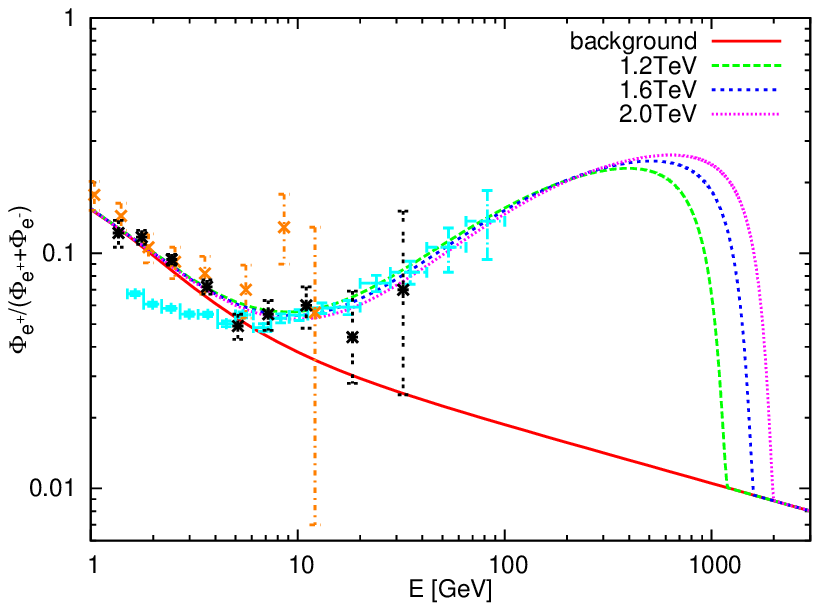}
\hspace*{5mm}
\epsfxsize=7cm
\leavevmode
\epsfbox{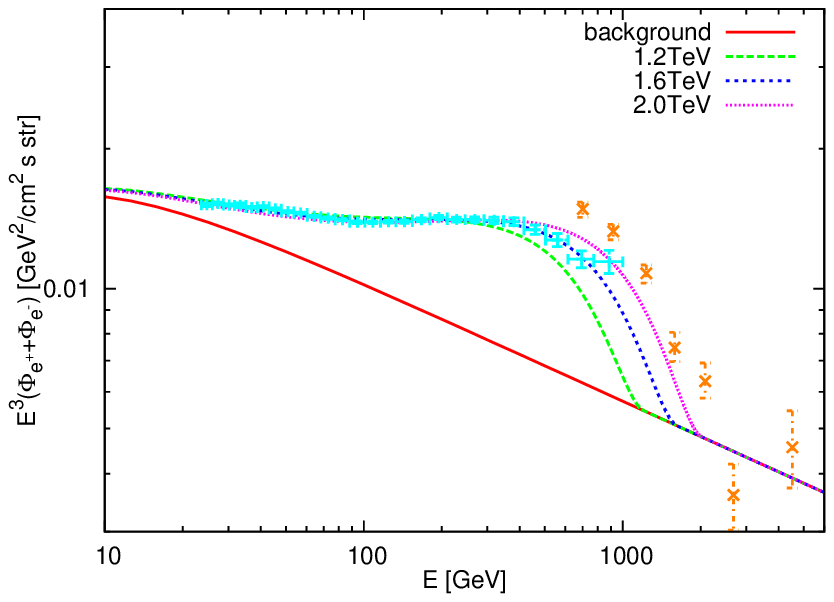}
\end{center}

{\footnotesize {\bf Fig.~4}~~Left and right panels show the predicted positron
 excess at the PAMELA regions and the prediction for
 the $e^++e^-$ flux at the observation regions of Fermi-LAT and H.E.S.S.,
 respectively. In both panels, DM mass $M_1$ and annihilation cross
 section $\langle\sigma_2v\rangle$ are fixed as 
$(M_1~(TeV),\langle\sigma_2v\rangle~({\rm cm}^3/{\rm sec}))=
(1.2, 4.1\times 10^{-23}),(1.6, 6.7\times 10^{-23}),
(2.0, 9.2\times 10^{-23})$. The normalization of background fluxes is
 taken to be $N_\Phi=0.64$.}
\end{figure}

Using these formulas, we plot the positron 
fraction $\Phi_{e^+}/(\Phi_{e^+}+\Phi_{e^-})$ and the total flux of
$e^++e^-$ scaled by $E^3$ in two panels of Fig.~4, respectively. 
In the left panel, the data points for positron excess 
of PAMELA \cite{pamela}, CAPRICE94 \cite{caprice94} and 
HEAT95 \cite{heat} are also plotted.
On the other hand, the data points for the $e^++e^-$ flux 
of Fermi-LAT \cite{fermi} and HESS \cite{hess} are plotted in the right panel. 
In this figure $\langle\sigma_2v\rangle$ is fixed to make the positron 
flux $\Phi_{e^+}$ to realize a good fit to the data of PAMELA and 
Fermi-LAT for each $M_1$ value. 
This figure shows that the flux of positrons and electrons 
predicted from the annihilation of $N_1$ in this extended model can give
rather good fits with these experimental data. 
Especially, the predicted flux fits well both data of the PAMELA and
Fermi-LAT experiments for $M_1=1.6$~TeV and 
$\langle\sigma_2v\rangle= 6.7\times 10^{-23}$~cm$^3$/sec.
This value of $\langle\sigma_2v\rangle$ can be realized by 
the parameter sets in case (ii) given in Table 1. 
If we apply this information for $M_1$ to
Fig.~1, we can predict the value of $Br(\mu\rightarrow e\gamma)$ and
$Br(\tau\rightarrow \mu\gamma)$.
The figure shows that the predicted value for $\mu\rightarrow e \gamma$
is within the reach of the MEG experiment \cite{meg}. 
Thus, lepton flavor violating processes could be
a crucial probe for this model. 

Annihilation of the DM can cause additional contributions to the cosmic gamma
ray. In fact, if hard charged leptons are produced as the final
states of the DM annihilation, high energy photons are also produced 
through several processes. 
One of their origin is the inverse Compton scattering of positrons
with CMB, star light and interstellar photon \cite{ic}. 
The other ones are final state radiation or internal radiation 
\cite{fsr}. 
The gamma ray flux expected from the former one does not depend on 
the particle physics model 
as long as the positron flux data presented by PAMELA is assumed. It can be 
used as a crucial constraint on the model. 
Since the gamma ray flux caused by the latter ones 
depends on the adopted particle physics model, the predicted photon 
spectrum can be used to discriminate the model from others on the basis
of the deviation of the photon spectrum from the one of background in
the future observation.

In the present scenario, the DM has mass of $O(1)$~TeV and it can decay
into $\tau^\pm$.
Thus, substantial constraints are expected to be imposed by the gamma from 
the former origin and also the gamma produced through the decay of
$\pi^0$ which comes from the $\tau^\pm$ decay. 
These give strong constraints on the gamma ray flux at higher energy 
regions. Various studies related to this issue have been done in the model 
independent way or in the fixed models \cite{ic,gamma2,gamma3}. 
The constraints obtained from analyses of the first year of Fermi
$\gamma$-ray observations are also given in \cite{gfermi}.
Their results for the gamma ray flux associated with the DM 
annihilation into charged lepton pairs are applicable
to our model to examine the consistency with
the diffuse gamma ray observations.
They show that the galactic diffuse gamma data 
constrain the assumed DM halo profile severely. Only the restricted halo 
profile called isothermal seems to be consistent 
with the observations.
In fact, following the study by Cirelli {\it et al.} in \cite{gfermi} 
for the cases with the final states $\mu^+\mu^-$ 
and $\tau^+\tau^-$ for the DM with mass around 1.6~TeV, 
the DM annihilation cross section $\langle\sigma v\rangle$ 
is shown to be less than $6\times 10^{-23}$~cm$^3$/sec 
and $1\times 10^{-22}$~cm$^3$/sec, respectively.
Papucci {\it et al.} in \cite{gfermi} gives much stronger constraints for
the $\tau^+\tau^-$ case. 
The results in this section show that our model can 
satisfy these constraints for $\mu^+\mu^-$ but the situation seems to be
marginal for the $\tau^+\tau^-$ case. 
Thus, the present model may be considered to work well in the isothermal
profile, although this type of halo profile is considered to be 
disfavored by the $N$-body simulation. 
We also note that the diffuse neutrino flux satisfies the present 
observational constraints \cite{ic,dneut}.  
 
\section{Summary}  

The radiative seesaw model is a simple and interesting extension 
of the SM by an inert doublet scalar and singlet fermions.
It can give the origin of both small neutrino masses and DM consistently. 
However, if we try to explain the positron excess observed by the PAMELA
experiment on the basis of the DM annihilation in this model,
an extremely large boost factor for the annihilation cross section is required. 
In this paper we have proposed a simple extension of the model by
introducing a singlet scalar. In this extended model,  
the DM annihilation cross section can be enhanced in 
the present Galaxy through the Breit-Wigner resonance 
without disturbing the features in the original model, which are favored 
by the neutrino masses, the lepton flavor violating processes 
and the DM relic abundance.
However, it should be noted that the mass of the
singlet scalar has to be finely tuned at the level of $O(10^{-6})$
for this enhancement.
Final states of DM annihilation are composed of heavier leptons only and
the ratio of $\mu^+$ and $\tau^\pm$ contribution to the annihilation
cross section is 1 to 3.
As a result of these features, the data for the positron and electron
flux observed by PAMELA and Fermi-LAT are well explained in this 
extended model as long as the coannihilation among $N_k$ occurs.
It is interesting that these results are closely related to the flavor
structure of neutrino Yukawa couplings, which induces tri-bimaximal mixing.

This extended model may be checked through the study of lepton flavor
violating processes such as $\mu\rightarrow e\gamma$ in the MEG
experiment and others in near future.
The cosmic positron and electron flux at higher energy regions 
may be clarified by the future CALET experiment,
which can observe $e^\pm$ flux up to 10 TeV \cite{calet}.
Viability of the model may also be confirmed through this experiment.
Although the diffuse gamma ray flux imposes severe constraints 
on the model, they could be consistent as long as the specific halo 
density profile called the isothermal profile is assumed. 
Detailed knowledge on the density profile of the DM halo seems
to be required to judge the validity of the explanation given here for 
the anomaly reported by PAMELA and Fermi-LAT.

\vspace*{5mm}
This work is partially supported by a Grant-in-Aid for Scientific
Research (C) from Japan Society for Promotion of Science (No.21540262).
Numerical computation was partially carried out by using the computing
facility at Yukawa Institute. 

\newpage
\noindent
{\Large\bf Appendix A}
\vspace*{3mm}

In order to study the stability of the
scalar potential, we need the renormalization group equations (RGEs) for the
coupling constants included in the scalar potential $V$.
This model is an extension of the SM with three singlet fermions $N_k$
and one doublet scalar $\eta$. Thus, the RGEs are similar to the ones 
of the ordinary two doublet Higgs model. However, we need to take
account of the effect of large neutrino Yukawa couplings $h_{\alpha k}$,
which are assumed to satisfy the relation (\ref{yukawa}).

If we assume the existence of the hidden sector gauge interaction mentioned
in the text, invariance of ${\cal L}_N$ restricts the representation of $N_k$ 
and $\eta$ to be an adjoint representation of SU($N$) or a doublet of
SU(2), for example. 
In case of the adjoint representation, scalar potential $V$ should be modified.
No anomaly problem appears in both cases.
However, if we note that one-loop diagrams with internal lines of $N_k$ and
$\eta$ have additional group theoretical factor dim($R$), we find that 
the latter is favored from the constraints of lepton flavor violating
processes. 

To prepare RGEs applicable to this extended situation, 
we assume that $N_k$ and $\eta$ are singlets ($N=1$) or doublets ($N=2$) 
of a hidden gauge symmetry SU(2). All SM fields are singlets under this 
group. 
A set of relevant RGEs can be written in the following form \cite{rge}:
\begin{eqnarray}
16\pi^2\frac{d\lambda_1}{dt}&=&24\lambda_1^2+2N\lambda_3^2+N\lambda_4^2
+2N\lambda_3\lambda_4+12\lambda_1h_t^2-6h_t^4+\kappa_\phi^2  ,\nonumber \\
16\pi^2\frac{d\lambda_2}{dt}&=&8(N+2)\lambda_2^2+2\lambda_3^2+\lambda_4^2
+2\lambda_3\lambda_4+4\lambda_2\left[2\left(h_{\tau 1}^2+h_{\tau
2}^2+\frac{3}{2}h_{\tau 3}^2\right)-3C_2(N)g_h^2\right] \nonumber \\
&&-8\left((h_{\tau 1}^2+h_{\tau 2}^2)^2+\frac{9}{4}
h_{\tau 3}^4\right)+\frac{3(N-1)(N^2+2N-2)}{4N^2}g_h^4 +\kappa_\eta^2,
\nonumber \\
16\pi^2\frac{d\lambda_3}{dt}&=&
 4\Big(3\lambda_1\lambda_3+(2N+1)\lambda_2\lambda_3+
 \lambda_1\lambda_4+N\lambda_2\lambda_4\Big)+4\lambda_3^2+2\lambda_4^2
\nonumber \\
&&+2\lambda_3\left[3h_t^2+2\left(h_{\tau 1}^2+h_{\tau 2}^2
+\frac{3}{2}h_{\tau
3}^2\right)-3C_2(N)g_h^2\right]+2\kappa_\phi\kappa_\eta,
\nonumber \\
16\pi^2\frac{d\lambda_4}{dt}&=& 
2\lambda_4\left[2\left(\lambda_1+\lambda_2+2\lambda_3+\lambda_4\right)
+3h_t^2+2\left(h_{\tau 1}^2+h_{\tau
		2}^2+\frac{3}{2}h_{\tau3}^2\right)
-3C_2(N)g_h^2\right], \nonumber \\
16\pi^2\frac{d\kappa}{dt}&=& 
20\kappa^2 +2\kappa_\phi^2 +2N\kappa_\eta^2 + 8N\sum_{k=1}^3(\kappa |y_k|^2
-4|y_k|^4), \nonumber \\
16\pi^2\frac{d\kappa_\phi}{dt}&=& 
12\kappa_\phi\lambda_1 +4N\kappa_\eta\lambda_3 
+2N\kappa_\eta\lambda_4
+8\kappa\kappa_\phi
+4\kappa_\phi^2 
+ 4N\kappa_\phi\sum_{k=1}^3 |y_k|^2+6\kappa_\phi h_t^2, \nonumber 
\end{eqnarray}
\begin{eqnarray}
16\pi^2\frac{d\kappa_\eta}{dt}&=& 
(8N+4)\kappa_\eta\lambda_2+4\kappa_\phi\lambda_3+4\kappa_\phi\lambda_4
+8\kappa\kappa_\eta+4\kappa_\eta^2  
+ 4\kappa_\eta\Big(h_{\tau 1}^2+h_{\tau
	       2}^2+\frac{3}{2}h_{\tau3}^2\Big) \nonumber\\
&&+4N\kappa_\eta\sum_{k=1}^3|y_k|^2 
-32\Big(h_{\tau 1}^2|y_1|^2+h_{\tau
	       2}^2|y_2|^2+\frac{3}{2}h_{\tau3}^2|y_3|^2\Big)
-6C_2(N)\kappa_\eta g_h^2, \nonumber \\
16\pi^2\frac{dh_{\tau i}}{dt}&=&h_{\tau i}
\Big((N+4)(h_{\tau 1}^2+h_{\tau
 2}^2)+3h_{\tau 3}^2+2|y_i|^2-3C_2(N)g_h^2 \Big)
\quad (i=1,2),\nonumber \\
16\pi^2\frac{dh_{\tau 3}}{dt}&=&h_{\tau 3}\Big(2(h_{\tau 1}^2+h_{\tau
 2}^2)+\frac{3}{2}(N+4)h_{\tau 3}^2+2|y_3|^2-3C_2(N)g_h^2\Big),
\nonumber \\
16\pi^2\frac{dy_i}{dt}&=&y_i\Big(8|y_i|^2+ 2N\sum_{k=1}^3|y_k|^2 +
2h_{\tau i}^2(2+\delta_{3i})\Big)  \quad (i=1,2,3), \nonumber \\ 
16\pi^2\frac{dh_t}{dt}&=&h_t\left(\frac{9}{2}h_t^2-8g_3^2\right),\nonumber \\
16\pi^2\frac{dg_h}{dt}&=&\frac{g_h^3}{3}\Big(-11N+\sum_{N_k,\eta}2T(N)\Big),
\nonumber \\
16\pi^2\frac{dg_3}{dt}&=&-7g_3^3,
\end{eqnarray}
where $C_2(n)$ and $T(n)$ stand for values of the second order Casimir 
operators defined by $\sum_aT^aT^a=C_2(n)1$ and 
${\rm tr}(T^aT^b)=T(n)\delta^{ab}$ for SU(n) generators $T^a$ 
in the fundamental representation. Thus, $g_h=0$ for the $N=1$ case, and  
$C_2(2)=\frac{3}{4}$ and $T(2)=\frac{1}{2}$ for the $N=2$ case, respectively.
In these RGEs we take account of the contributions to 
$\beta$-functions only from the top Yukawa coupling $h_t$, 
the neutrino Yukawa coupling $h_{\alpha k}$, the strong gauge 
coupling $g_3$ and the hidden gauge coupling $g_h$ except for the 
couplings in the scalar potential. 
Since the $\beta$-function of $\lambda_5$ is 
proportional to $\lambda_5$ due to the symmetry discussed in the text,
it is kept sufficiently small to be neglected. 
\vspace*{4mm}

\noindent
{\Large\bf Appendix B}
\vspace*{3mm}

\noindent
In the present model the final state positron is yielded as a
consequence 
of $\mu^+$ and $\tau^\pm$ decay.
We determine the energy spectrum of such positrons by using the 
PYTHIA Monte Carlo code \cite{code}.  
If we write an expectation value of the number of this yielded positron 
per the decay of $\alpha(=\mu^+, \tau^\pm)$ as $N_{\alpha, e^+}$, 
PYTHIA gives the positron spectrum $\displaystyle \frac{dN_{\alpha,e^+}}{dE}$. 
The spectrum obtained from this simulation is shown in Fig.~5, where the result 
for the $\alpha=\mu^+$ is plotted in the left panel and
the one for $\alpha=\tau^\pm$ pair is plotted in the right panel. 
We find from these figures that the positron produced 
through the decay of $\tau^\pm$ is softer than the one for
$\mu^+$ as mentioned in the text.  

\begin{figure}[ht]
\begin{center}
\epsfxsize=7cm
\leavevmode
\epsfbox{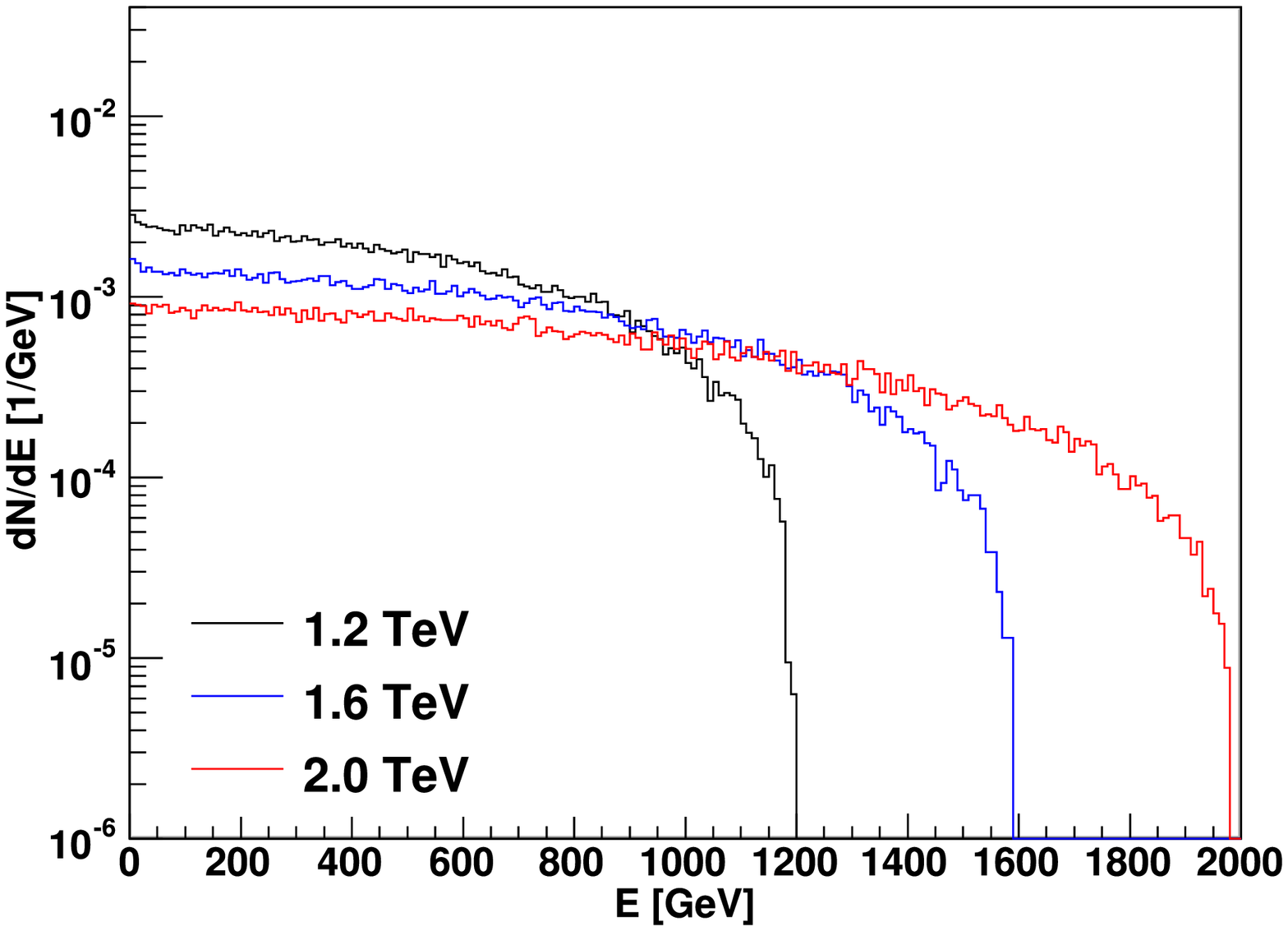}
\hspace*{5mm}
\epsfxsize=7cm
\leavevmode
\epsfbox{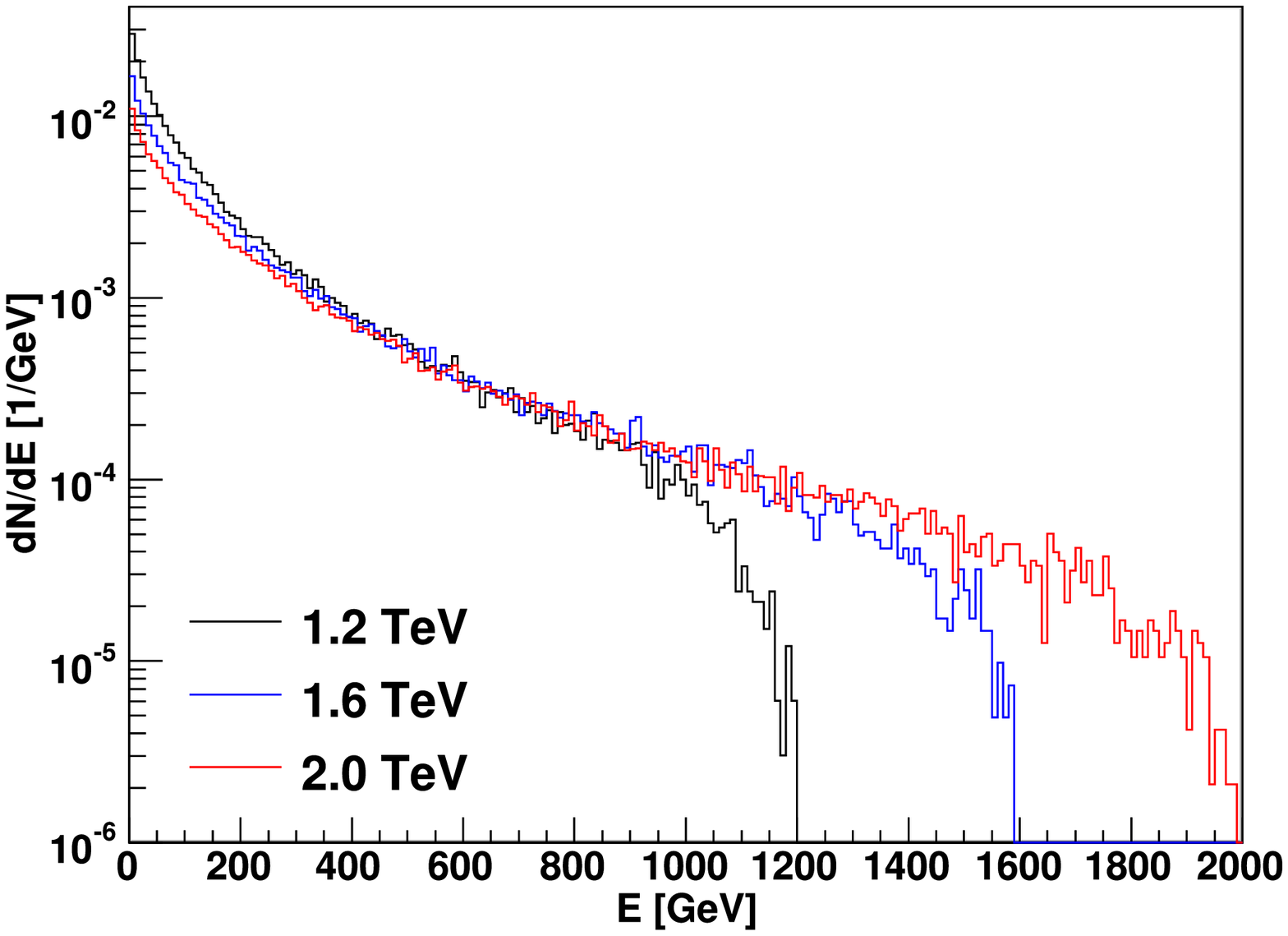}
\end{center}

{\footnotesize {\bf Fig.~5}~~The energy spectrum 
$\displaystyle \frac{dN_{\alpha,e^+}}{dE}$ obtained for
the DM mass $M_1=1.2,~1.6,~2.0$~TeV by simulation.
Left and right panels show the positron spectrum obtained from 
the decay of the $\mu^+$ and $\tau^\pm$ pair, respectively.} 
\end{figure}

In order to fix their empirical formulas approximately, 
each data set in Fig.~5 are fitted by using the functions
\begin{equation}
\frac{dN_{\alpha,e^+}}{dE}=\sum_{n=0}^2 \frac{d_{n}(M_1-E)^{1/2}}
{(E+ E_0)^n} 
\end{equation}
where $E_0$ is a constant and $E$ should be understood in a GeV unit.
As results of this fitting, we find that the coefficients $d_{n}$ 
in the above fitting functions should take the values shown in Table 2. 
We have $N_{\mu^+,e^+}=1$, $N_{\tau^\pm,e^+}\sim 1.3$ 
by integrating the obtained spectrum.
This corresponds to the fact that the decay of $\tau^+$ is composed of 
various modes such as $\tau^+\rightarrow e^+\bar{\nu}_{\tau}\nu_e$, 
$\tau^\pm\rightarrow \mathrm{hadrons}\rightarrow e^\pm e^\pm e^\mp$,
while the decay mode is dominated only by 
$\mu^+\rightarrow e^+\bar{\nu}_\mu\nu_e$ for $\mu^+$. 

\scriptsize
\begin{center}
\begin{tabular}{c|ccc|ccc}\hline
particle($\alpha$) &&$\mu^+$&&&$\tau^\pm$&\\ \hline
$M_1$(TeV) & 1.2 & 1.6 & 2.0 & 1.2 & 1.6 & 2.0 \\ \hline
$d_0$ & $-2.76\times 10^{-3}$ & $-3.86\times 10^{-3}$ &
 $-2.13\times 10^{-3}$ & $-2.50\times 10^{-6}$ &
 $-1.14\times 10^{-6}$ & $-7.75\times 10^{-7}$ \\
$d_1$ & $4.13\times 10^{1}$ & $1.20\times 10^{2}$ &
 $7.19\times 10^{1}$ & $2.94\times 10^{-3}$ &
 $1.85\times 10^{-3}$ & $1.44\times 10^{-3}$ \\
$d_2$ & $-1.52\times 10^{5}$ & $-9.26\times 10^{5}$ &
 $-6.01\times 10^{5}$ & $2.43\times 10^{0}$ &
 $3.27\times 10^{0}$ & $3.86\times 10^{0}$ \\
$E_0$ & $7.12\times 10^{3}$ & $1.51\times 10^{4}$ &
 $1.64\times 10^{4}$ & $6.54\times 10^{1}$ &
 $9.60\times 10^{1}$ & $1.24\times 10^{2}$ \\\hline
\end{tabular}
\vspace*{5mm}\\
{\footnotesize {\bf Table.~2}~~The coefficients $d_n$ determined by the
 fitting to Fig.~5. }
\end{center}
\normalsize

In the text we make the estimation of the positron flux (\ref{pflux2}) 
by using these positron spectra.
The parameters included in the cross section 
$\langle\sigma_2v\rangle$ is determined by a point 
in the allowed regions (on the red solid line) shown in Fig.~1. 
Other parameters $\Delta$ and $y_k$ relevant only 
to the $N_1$ annihilation at the present Galaxy are  
fixed to make $\langle\sigma_2v\rangle$ a suitable
value $O(10^{-23})$~cm$^3$/sec for the explanation of PAMELA data.    
They are summarized in Table 1 in the case of $M_1=1.6$~TeV.\\

\end{document}